\documentclass[twocolumn,pra,superscriptaddress,showpacs]{revtex4}
\usepackage[utf8]{inputenc}
\usepackage{amsmath}
\usepackage{amssymb}
\usepackage{graphicx}
\usepackage{esint}
\usepackage[unicode=true,pdfusetitle,
 bookmarks=true,bookmarksnumbered=false,bookmarksopen=false,
 breaklinks=false,pdfborder={0 0 1},backref=section,colorlinks=false]
 {hyperref}
\hypersetup{
 colorlinks,citecolor=blue}
\usepackage{breakurl}

\makeatletter
\@ifundefined{textcolor}{}
{%
 \definecolor{BLACK}{gray}{0}
 \definecolor{WHITE}{gray}{1}
 \definecolor{RED}{rgb}{1,0,0}
 \definecolor{GREEN}{rgb}{0,1,0}
 \definecolor{BLUE}{rgb}{0,0,1}
 \definecolor{CYAN}{cmyk}{1,0,0,0}
 \definecolor{MAGENTA}{cmyk}{0,1,0,0}
 \definecolor{YELLOW}{cmyk}{0,0,1,0}
}

\usepackage{subfigure}\usepackage{epsfig}\usepackage{amsfonts}\usepackage{mathrsfs}\usepackage[toc,page,title,titletoc,header]{appendix}

\makeatother

\begin{document}

\title{Quantum Defect Theory for Orbital Feshbach Resonance }

\author{Yanting Cheng}

\affiliation{Institute for Advanced Study, Tsinghua University, Beijing, 100084,
China}

\author{Ren Zhang}
\email{rine.zhang@gmail.com}


\affiliation{Institute for Advanced Study, Tsinghua University, Beijing, 100084,
China}

\author{Peng Zhang}
\email{pengzhang@ruc.edu.cn}


\affiliation{Department of Physics, Renmin University of China, Beijing, 100872,
China}

\affiliation{Beijing Computational Science Research Center, Beijing, 100084, China}

\affiliation{Beijing Key Laboratory of Opto-electronic Functional Materials \&
Micro-nano Devices (Renmin University of China)}
\begin{abstract}
In the ultracold gases of alkali-earth (like) atoms, a new type of
Feshbach resonance, i.e., the orbital Feshbach resonance (OFR), has
been proposed and experimentally observed in ultracold $^{173}$Yb
atoms. When the OFR of the $^{173}$Yb atoms occurs, the energy gap
between the open and closed channels is smaller by two orders of magnitudes
than the van der Waals energy. As a result, quantitative
accurate results for the low-energy two-body problems can be obtained
via multi-channel quantum defect theory (MQDT), which is based on
the exact solution of the Schr$\ddot{{\rm o}}$dinger equation with
the van der Waals potential. In this paper we use the MQDT to calculate
the two-atom scattering length, effective range, and the binding energy
of two-body bound states for the systems with OFR. With these results
we further study the clock-transition spectrum for the two-body bound states,
which can be used to experimentally measure the binding energy. Our
results are helpful for the quantitative theoretical and experimental
researches for the ultracold gases of alkali-earth (like) atoms with
OFR. 
\end{abstract}

\maketitle

\section{introduction}

Feshbach resonance \cite{fr} is a powerful tool for the control of
interaction between ultracold atoms \cite{chinrmp}. In ultracold
gases of alkali atoms the magnetic Feshbach resonances are widely
used for the tuning of $s$-wave scattering lengths \cite{chinrmp}.
For the gases of ultracold alkali-earth (like) atoms, recently we
found a new type of Feshbach resonance, i.e., the orbital Feshbach
resonance (OFR) \cite{ourprl}. With the help of OFR, one can precisely
control the $s$-wave scattering length between two fermionic alkali-earth
(like) atoms in $^{1}{\rm S}_{0}$ and $^{3}{\rm P}_{0}$ electronic
orbital states with different nuclear spin, by changing the magnetic
field \cite{ourprl}. OFR has been experimentally observed in the
ultracold gases of $^{173}$Yb atoms \cite{ofrexp1,ofrexp2}. It is
also shown that using the ultracold gases of alkali-earth (like) atoms
with OFR one can study several interesting problems, including the
Kondo effect, the strong-interacting ultracold fermi gases with narrow
Feshbach resonance and the Leggett mode, \textit{et. al.} \cite{ourprl,kondo,nr,c1,c2,c3,c4,c5,c6,c7}.

When the OFR of $^{173}$Yb atoms occurs, the energy gap between the
open and the closed channel is about $10^5$Hz. It is by two orders of magnitude smaller
than the characteristic energy of the inter-atom interaction (i.e.,
the van der Waals energy), which is of the order of $10^7$Hz \cite{ofrexp1,ofrexp2,spinexchange3,ourpra}.
As a result, a simple zero-range two-channel Huang-Yang pseudopotential
can be used as an approximation for the inter-atom interaction \cite{ourprl,ourpra}.
In this model, the two-body interaction is described by
two parameters, i.e., the scattering lengths $a_{\pm}$ for the two
independent scattering channels $|\pm\rangle$ which will be defined
below. It is estimated that for $^{173}$Yb atoms the quantitative
precision of the OFR point given by the two-channel Huang-Yang pseudopotential
is about $80\%$ \cite{ourprl,ourpra}.

To obtain more accurate results, one needs to take into account the
effects from the finite range van der Waals interaction potential.
To this end, one can numerically solve the multi-channel Schr$\ddot{{\rm o}}$dinger
equation with a model interaction potential which behaves as a van
der Waals potential in the long-range limit (\textit{e.g.,} the Lenard-Jones
potential) \cite{c5}. Nevertheless, there is also an analytical approach
for the multi-channel low-energy two-body problem with van der Waals
potential, i.e., the multi-channel quantum defect theory (MQDT) \cite{qdt1,qdt2,qdt3,qdt4}
which is based on the analytical solution of the single-channel Schr$\ddot{{\rm o}}$dinger
equation with van der Waals potential \cite{van}. In ultracold atomic
gas physics, this MQDT approach was originally developed for alkali
atoms. Previous research for these systems shows that when the inter-channel
energy gap is much smaller than the van der Waals energy, the result
given by the MQDT is quantitatively very accurate \cite{qdt2,qdt4}.
Thus, this approach is also applicable for the ultracold alkali earth
(like) atoms with an OFR with small energy gaps between the open and
the closed channels, \textit{e.g.}, the ultracold $^{173}$Yb atoms.

In this paper, using the MQDT we solve the low-energy two-body problems for alkali earth
(like) atoms with an OFR. We derive the analytical expressions of the
two-atom scattering length and effective range (Eqs. (\ref{as}) and
(\ref{asreff})), as well as the algebraic equation satisfied by the
binding energy of two-body bound state (Eq. (\ref{Eb})). All the
results are expressed in terms of the scattering length $a_{\pm}$
as well as the characteristic length $\beta_{6}$ of the van der Waals
potential. Our results show that the OFR for $^{173}{\rm Yb}$ atoms
is a narrow resonance\cite{nr}. Using these results we further investigate
the clock-transition spectrum of these systems, which can be used
for the experimental measurement of the binding energy. Our results
are helpful for both theoretical and experimetnal study for ultracold
alkali-earth (like) atoms with OFR.

The remainder of this manuscript is organized as follows. In Sec.
II we show the MQDT approach for our system and calculate the two-atom
$s$-wave scattering length and effective range. In Sec. III we calculate
the binding energy and wave function of the two-atom bound state,
as well as the clock-transition spectrum. Some summaries and discussions
for our results are presneted in Sec. IV, and
some details of our calculations are  shown in the appendixes.

\section{scattering length and effective range}

\begin{figure}
\includegraphics[width=6.5cm]{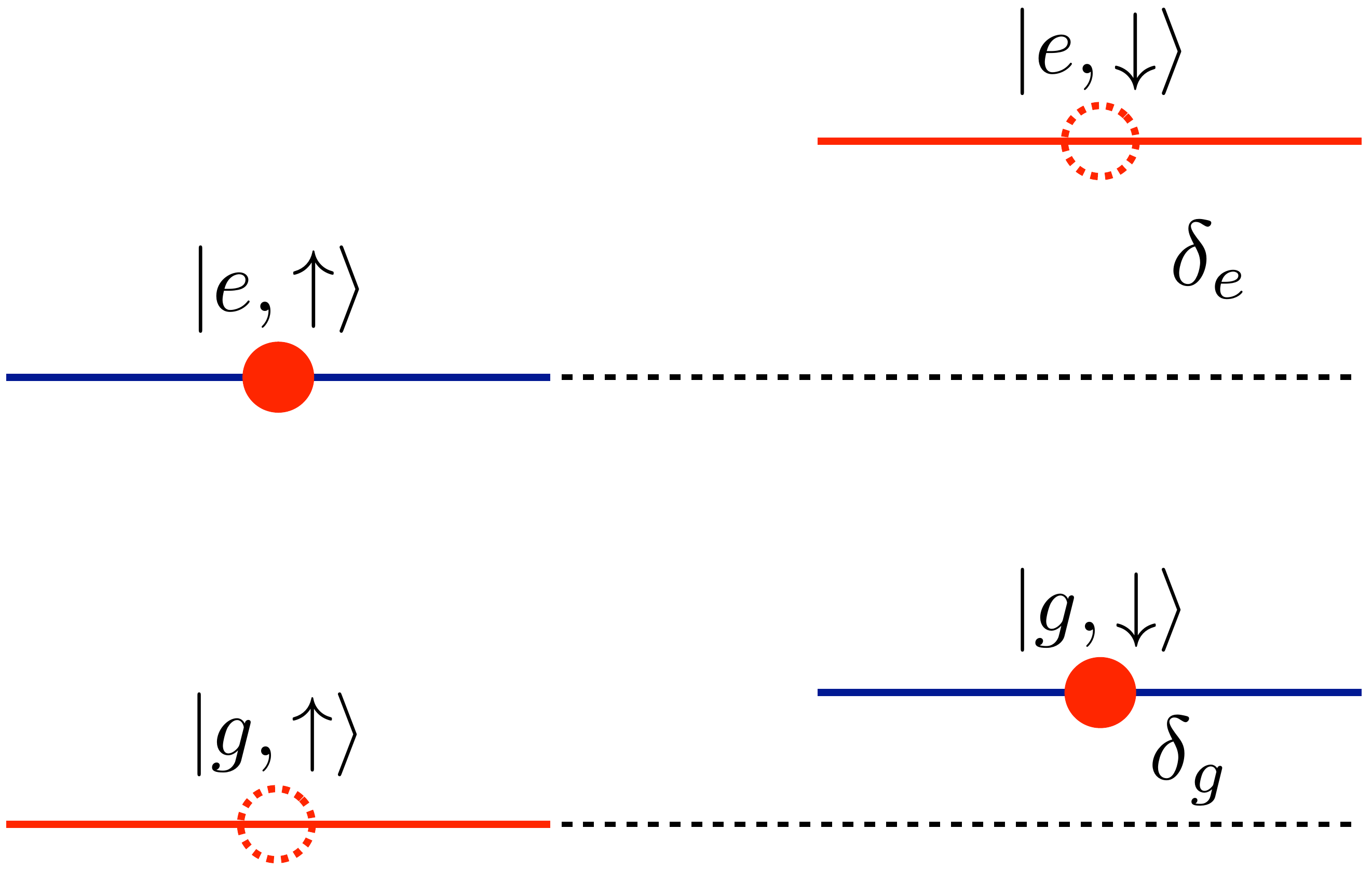} \caption{(color online) Energy level diagram of a single atom. For the OFR,
the open channel $|o\rangle$ is the two-body internal state with
one atom being in $|g,\downarrow\rangle$ and the other being in $|e,\uparrow\rangle$
(the red filled circles), while the closed channel $|c\rangle$ is
the state where one atom being in $|g,\uparrow\rangle$ and the other
being in $|e,\downarrow\rangle$ (the red unfilled circles with dashed
line). Here $\delta_{e}=\mu_{e}\Delta_{m}B$ is the Zeeman-energy
difference between the single-atom states $|e,\downarrow\rangle$
and $|e,\uparrow\rangle$, and $\delta_{g}=\mu_{g}\Delta_{m}B$ is
the one between states $|g,\downarrow\rangle$ and and $|g,\uparrow\rangle$,
with $\mu_{e(g)}$ being the magnetic moment for the electronic orbital
state $|e\rangle$ ($|g\rangle$). The Zeeman energy difference $\delta$
in Eq. (\ref{h}) can be expressed as $\delta=\delta_{e}-\delta_{g}=(\delta\mu)B$
with $\delta\mu=(\mu_{e}-\mu_{g})\Delta_{m}$. \label{fig:energy-level}}
\end{figure}

We consider two fermionic alkali-earth (like) atoms in $^{1}{\rm S}_{0}$
($g$) and $^{3}{\rm P}_{0}$ ($e$) electronic orbital states, with
nuclear-spin magnetic quantum numbers $m_{I}$ ($\uparrow$) and $m_{I}+\Delta_{m}$
($\downarrow$) (Fig. \ref{fig:energy-level}). The two-body internal
state with one atom being in $|g,\downarrow\rangle$ and the other
being in $|e,\uparrow\rangle$ can be denoted as 
\begin{equation}
|o\rangle\equiv|g,\downarrow;e,\uparrow\rangle.\label{o}
\end{equation}
Similarly, we also denote the state with one atom being in $|g,\uparrow\rangle$
and the other being in $|e,\downarrow\rangle$ as 
\begin{equation}
|c\rangle\equiv|g,\uparrow;e,\downarrow\rangle.\label{c}
\end{equation}

The Hamiltonian for the two-atom relative motion is given by 
\begin{equation}
\hat{H}=-\frac{\hbar^{2}}{m}\nabla_{{\bf r}}^{2}+\delta|c\rangle\langle c|+U(r),\label{h}
\end{equation}
where $m$ is the single-atom mass, ${\bf r}$ is the relative position
of these two atoms and $\delta=(\delta\mu)B$ is the Zeeman-energy
difference between states $|c\rangle$ and $|o\rangle$, with $\delta\mu$
and $B$ being the magnetic moment difference of these two states
and the magnetic field, respectively. Without loss of generality,
here we assume that $\delta\mu>0$. In Eq. (\ref{h}) $U(r)$
is the inter-atom interaction potential. It is diagonal in the bases
\begin{equation}
|\pm\rangle=\frac{1}{\sqrt{2}}\left(|c\rangle\mp|o\rangle\right),\label{p-1}
\end{equation}
and can be expressed as 
\begin{equation}
U(r)=U^{(+)}(r)|+\rangle\langle+|+U^{(-)}(r)|-\rangle\langle-|,\label{ur}
\end{equation}
where $U^{(\pm)}(r)$ is the potential curve with respect to state
$|\pm\rangle$. When the two atoms are far away enough from each other,
$U^{(\pm)}(r)$ can be approximated as the same van der Waals potential,
i.e, we have 
\begin{equation}
U^{(+)}(r>b)\approx U^{(-)}(r>b)\approx-\frac{\hbar^{2}\beta_{6}^{4}}{mr^{6}}.\label{uvdw}
\end{equation}
Here $\beta_{6}$ is the characteristic length of the van der Waals
potential and the range $b$ satisfies the conditions 
\begin{eqnarray}
b & < & \beta_{6};\label{b}\\
\frac{\hbar^{2}\beta_{6}^{4}}{mb^{6}} & \gg & \frac{\hbar^{2}}{m\beta_{6}^{2}}.\label{b2}
\end{eqnarray}
In this paper we focus on the systems where the energy gap $\delta$
between the states $|c\rangle$ and $|o\rangle$ is much smaller than
the van der Waals energy $\hbar^{2}/(m\beta_{6}^{2})$. As shown below,
our finial result is independent of the exact value of $b$.

We consider the $s$-wave scattering of two atoms incident from channel
$|o\rangle$, with relative momentum $\hbar k$. Here we assume the
scattering energy 
\begin{equation}
\epsilon=\frac{\hbar^{2}k^{2}}{m}\label{esca}
\end{equation}
is smaller than the inter-channel energy gap $\delta$. As a result,
in the scattering process the channel $|o\rangle$ is open while the
channel $|c\rangle$ is closed.

The $s$-wave scattering length and effective range is determined
by the two-atom scattering wave function $|\psi_{\epsilon,\delta}(r)\rangle$
which satisfies the Schr$\ddot{{\rm o}}$dinger equaiton 
\begin{equation}
\hat{H}|\psi_{\epsilon,\delta}(r)\rangle=E|\psi_{\epsilon,\delta}(r)\rangle\label{se}
\end{equation}
with the boundary conditions 
\begin{equation}
\lim_{r\rightarrow0}\left[r\langle r|\psi_{\epsilon,\delta}(r)\rangle\right]=0\label{bc}
\end{equation}
and 
\begin{equation}
\lim_{r\rightarrow\infty}\langle c|\psi_{\epsilon,\delta}(r)\rangle\rightarrow0.\label{bc2-1}
\end{equation}
It is pointed out that, if we solve Eq. (\ref{se}) only with the
the boundary condition (\ref{bc}), we can get \textit{two} linearly-independent
special solutions. The solution of Eq. (\ref{se}) and conditions
(\ref{bc}, \ref{bc2-1}), i.e., the scattering wave function $|\psi_{\epsilon,\delta}(r)\rangle$,
can be expressed as the superposition of these two special solutions.
Following the idea of MQDT, below we will first derive the special
solutions of Eq. (\ref{se}) and the condition (\ref{bc}) for the
simple case with $\delta=0$, and then derive the special solutions
of (\ref{se}) and (\ref{bc}) for non-zero $\delta$. Finally, we
will construct the scattering wave function $|\psi_{\epsilon,\delta}(r)\rangle$
with these special solutions and the condition (\ref{bc2-1}). With
this wave function we can derive the $s$-wave scattering phase shift,
scattering length and effective range.

\subsection{Special Solutions of Eqs. (\ref{se}) and (\ref{bc}) for $\delta=0$}

When $\delta=0$, the Hamiltonian $\hat{H}$ given by Eq. (\ref{h})
is diagonal in the bases $\left\{ |+\rangle,|-\rangle\right\} $.
Therefore, in this case we can choose the two special solutions of
Eq. (\ref{se}) and (\ref{bc}) as 
\begin{eqnarray}
|\psi_{\epsilon,\delta=0}^{(+)}(r)\rangle & = & \frac{\phi_{\epsilon}^{(+)}(r)}{r}|+\rangle;\label{psid0p}\\
|\psi_{\epsilon,\delta=0}^{(-)}(r)\rangle & = & \frac{\phi_{\epsilon}^{(-)}(r)}{r}|-\rangle.\label{psid0m}
\end{eqnarray}
Substituting Eqs. (\ref{psid0p}, \ref{psid0m}) into Eq. (\ref{se}),
we obtain two equaitons for the components $\phi_{\epsilon}^{(\pm)}(r)$:
\begin{equation}
-\frac{\hbar^{2}}{m}\frac{d^{2}}{dr^{2}}\phi_{\epsilon}^{(\pm)}(r)+U^{(\pm)}(r)\phi_{\epsilon}^{(\pm)}(r)=\epsilon\phi_{\epsilon}^{(\pm)}(r).\label{ed0}
\end{equation}
Furthermore, using the epression (\ref{uvdw}) of the potential $U^{(\pm)}(r)$
in the region $r>b$, in this region we can reduce these two equations
to 
\begin{equation}
-\frac{d^{2}}{dr^{2}}\phi_{\epsilon}^{(\pm)}(r)-\frac{\beta_{6}^{4}}{r^{6}}\phi_{\epsilon}^{(\pm)}(r)=k^{2}\phi_{\epsilon}^{(\pm)}(r).\label{evdw}
\end{equation}
Thus, when $r>b$ the components $\phi_{\epsilon}^{(\pm)}(r)$ can
be expressed as the superpositions of two special solutions $f_{\epsilon}^{0}(r)$
and $g_{\epsilon}^{0}(r)$ of Eq. (\ref{evdw}), which were analytically
derived by Bo Gao in Ref. \cite{van}. Namely, we can choose $\phi_{\epsilon}^{(\pm)}(r)$
to satisfy 
\begin{eqnarray}
\phi_{\epsilon}^{(\pm)}(r>b) & = & f_{\epsilon}^{0}(r)-K_{\pm}^{0}g_{\epsilon}^{0}(r),\label{psip}
\end{eqnarray}
where the parameters $K_{\pm}^{0}$ are determined by the short-range
detail of the potential curves $U^{(\pm)}(r)$ in the region $r<b$.
In this paper we consider the case where the scattering energy $\epsilon$
is much smaller than the van der Waals energy $\hbar^{2}/(m\beta_{6}^{2})$.
Due to the conditions (\ref{b}) and (\ref{b2}), $\epsilon$ is also
much smaller than $\hbar^{2}/(mb^{2})$. In this case the behavior of
$\phi_{\epsilon}^{(\pm)}(r)$ in the region around $r=b$ is almost
independent of $\epsilon$. Furthermore, it is also shown that the
functions $f_{\epsilon}^{0}(r)$ and $g_{\epsilon}^{0}(r)$ are almost
$\epsilon$-independent when $r\approx b$ \cite{van}. Thus, the
parameters $K_{\pm}$ is alsom almost independent of $\epsilon$.
That is one of the basic ideas of the QDT \cite{qdt1,qdt5,qdt6}.

For $\epsilon>0$, in the limit $r\rightarrow\infty$ we have \cite{van}
\begin{eqnarray}
\lim_{r\rightarrow\infty}f_{\epsilon}^{0}(r) & = & \sqrt{\frac{2}{\pi k}}\left[Z_{ff}(\epsilon)\sin\left(kr\right)-Z_{fg}(\epsilon)\cos\left(kr\right)\right];\nonumber \\
\label{ff}\\
\lim_{r\rightarrow\infty}g_{\epsilon}^{0}(r) & = & \sqrt{\frac{2}{\pi k}}\left[Z_{gf}(\epsilon)\sin\left(kr\right)-Z_{gg}(\epsilon)\cos\left(kr\right)\right],\nonumber \\
\label{gg}
\end{eqnarray}
where 
\begin{equation}
k=\frac{\hbar}{\sqrt{m\epsilon}},\label{kk}
\end{equation}
and the functions $Z_{ij}(\epsilon)$ ($i,j=f,g$) are given in Ref.
\cite{van}. Substituting Eq. (\ref{ff}, \ref{gg}) into Eqs. (\ref{psip})
and using the expressions of s $Z_{ij}(\epsilon)$, one can obtain \cite{qdt1}
\begin{equation}
\lim_{r\rightarrow\infty}\phi_{\epsilon=0}^{(\pm)}(r)\propto\left[r-\frac{(2\pi)(K_{\pm}^{0}-1)}{\Gamma\left(1/4\right)^{2}K_{\pm}^{0}}\beta_{6}\right].\label{xx}
\end{equation}
On the other hand, since in our case with $\delta=0$ the states $|+\rangle$
and $|-\rangle$ are two independent scattering channels, we also
have $\lim_{r\rightarrow\infty}\phi_{\epsilon=0}^{(\pm)}(r)\propto(r-a_{{\rm s}}^{(\pm)})$
where $a_{{\rm s}}^{(\pm)}$ is the $s$-wave scattering length for
each channel. Thus, Eq. (\ref{xx}) implies the relation between
parameter $K_{\pm}^{0}$ and the scattering length $a_{{\rm s}}^{(\pm)}$  \cite{qdt1}:
\begin{equation}
K_{\pm}^{0}=\frac{2\pi\beta_{6}}{2\pi\beta_{6}-a_{{\rm s}}^{(\pm)}\Gamma\left(1/4\right)^{2}}.\label{kpm}
\end{equation}

\subsection{Special Solutions of Eqs. (\ref{se}) and (\ref{bc}) for $\delta\protect\neq0$}

Now we consider the special solutions of Eq. (\ref{se}) and condition
(\ref{bc}) for the case with $\delta\neq0$. As mentioned above,
in this subsection we ignore the boundary condition (\ref{bc2-1}).
When $\delta\neq0$, it is convenient to expand $|\psi_{\epsilon,\delta}(r)\rangle$
in the bases $\left\{ |c\rangle,|o\rangle\right\} $. Since 
the potential $U^{(\pm)}(r)$ satisfies the condition
(\ref{uvdw}), in the
region $r>b$  the interaction $U$ is independent
on the internal state of these two atoms. As a result, the channels
$|c\rangle$ and $|o\rangle$ are decoupled and Eq. (\ref{se}) can
be simplified as 
\begin{eqnarray}
\left[-\frac{d^{2}}{dr^{2}}-\frac{\beta_{6}^{4}}{r^{6}}\right]\left[r\langle o|\psi_{\epsilon,\delta}(r)\rangle\right] & = & \epsilon\left[r\langle o|\psi_{\epsilon,\delta}(r)\rangle\right];\label{eo}\\
\left[-\frac{d^{2}}{dr^{2}}-\frac{\beta_{6}^{4}}{r^{6}}\right]\left[r\langle c|\psi_{\epsilon,\delta}(r)\rangle\right] & = & (\epsilon-\delta)\left[r\langle c|\psi_{\epsilon,\delta}(r)\rangle\right].\nonumber \\
\label{ec}
\end{eqnarray}
Therefore, similar as in Sec. II. A, for $r>b$ the component
$r\langle o|\psi_{\epsilon,\delta}(r)\rangle$ can be expressed as
the superpositions of functions $f_{\epsilon}^{0}(r)$ and $g_{\epsilon}^{0}(r)$, and
$r\langle c|\psi_{\epsilon,\delta}(r)\rangle$can
be expressed as the superpositions of $f_{\epsilon-\delta}^{0}(r)$
and $g_{\epsilon-\delta}^{0}(r)$. Thus, we can choose the two special
solutions of $|\psi_{\epsilon,\delta}^{(\alpha,\beta)}(r)\rangle$
of Eq. (\ref{se}) to satisfy 
\begin{eqnarray}
&&|\psi_{\epsilon,\delta}^{(\alpha)}(r>b)\rangle\nonumber\\
&=& \frac{1}{r}\{
\left[f_{\epsilon}^{0}(r)-K_{oo}^{0}g_{\epsilon}^{0}(r)\right]|o\rangle-K_{co}^{0}g_{\epsilon-\delta}^{0}(r)|c\rangle\},
\label{psia}\\
&&|\psi_{\epsilon,\delta}^{(\beta)}(r>b)\rangle\nonumber\\
 & = &\frac{1}{r}\{ -K_{oc}^{0}g_{\epsilon}^{0}(r)|o\rangle+\left[f_{\epsilon-\delta}^{0}(r)-K_{cc}^{0}g_{\epsilon-\delta}^{0}(r)\right]|c\rangle\}.
 \nonumber \\
\label{psib}
\end{eqnarray}
Here the parameter $K_{ij}^{0}$ ($i,j=o,c$) is also determined by
the detail of the potential curves $U^{(\pm)}(r)$ in the region $r<b$.
Similar as in Sec. II. A, due to the conditions (\ref{b}, \ref{b2})
and the fact that both $\epsilon$ and $\delta$ are much smaller
than $\hbar^{2}/(m\beta_{6}^{2})$, the values of $K_{ij}^{0}$ ($i,j=o,c$)
is independent of both the scattering energy $\epsilon$ and the energy
gap $\delta$ \cite{qdt2} (Appendix A). Therefore, we can obtain
the values of $K_{ij}^{0}$ ($i,j=o,c$) from the behaivor of $|\psi_{\epsilon,\delta}^{(\alpha,\beta)}(r)\rangle$
in the limit $\delta\rightarrow0$ with the following analysis:
in Sec. II. A we have already obtained two special
solutions $|\psi_{\epsilon,\delta=0}^{(\pm)}(r)\rangle$ for Eqs. (\ref{se})
and (\ref{bc}) with $\delta=0$. Therefore, $|\psi_{\epsilon,\delta=0}^{(\alpha,\beta)}(r)\rangle$
should be the linear combinations of these two solutions,
i.e., $|\psi_{\epsilon,\delta}^{(\alpha,\beta)}(r)\rangle$ can be
expressed as 
\begin{eqnarray}
|\psi_{\epsilon,\delta=0}^{(\alpha)}(r>b)\rangle & = & A_{1}|\psi_{\epsilon,\delta=0}^{(+)}(r>b)\rangle+A_{2}|\psi_{\epsilon,\delta=0}^{(-)}(r>b)\rangle;\nonumber \\
\label{p1}\\
|\psi_{\epsilon,\delta=0}^{(\beta)}(r>b)\rangle & = & A_{3}|\psi_{\epsilon,\delta=0}^{(+)}(r>b)\rangle+A_{4}|\psi_{\epsilon,\delta=0}^{(-)}(r>b)\rangle,\nonumber \\
\label{p2}
\end{eqnarray}
with $A_{1,2,3,4}$ being $r$-independent coefficient. 
Substituting Eqs. (\ref{psia}, \ref{psib}) into the left-hand side of Eqs. (\ref{p1}, \ref{p2}), and
substituting Eqs. (\ref{psid0p}, \ref{psid0m}, \ref{psip}) 
into the ringt-hand side of Eqs. (\ref{p1}, \ref{p2})
 we find that
 \begin{eqnarray}
K_{cc}^{0} & = & K_{oo}^{0}=\frac{1}{2}\left(K_{+}^{0}+K_{-}^{0}\right);\label{kcc}\\
K_{co}^{0} & = & K_{oc}^{0}=\frac{1}{2}\left(K_{-}^{0}-K_{+}^{0}\right).\label{koc}
\end{eqnarray}
Moreover, with the relation (\ref{kpm}) the parameters $K_{ij}^{0}$
($i,j=o,c$) can be further expressed as functions of $\beta_{6}$
and the scattering length $a_{{\rm s}}^{(\pm)}$.

\subsection{Scattering Wave Function and Phase Shift}

Now we calculate the scattering wave function $|\psi_{\epsilon,\delta}(r)\rangle$
which satisfies Eq. (\ref{se}) and both of the two boundary conditions
(\ref{bc}) and (\ref{bc2-1}). This scattering state is the superposition
of the two special solutions $|\psi_{\epsilon,\delta}^{(\alpha,\beta)}(r)\rangle$
of Eqs. (\ref{se}) and (\ref{bc}), which were derived in Sec. II.
B. Namely, $|\psi_{\epsilon,\delta}(r)\rangle$ can be expressed as
\begin{equation}
|\psi_{\epsilon,\delta}(r)\rangle=B\left\{ |\psi_{\epsilon,\delta}^{(\alpha)}(r)\rangle+C|\psi_{\epsilon,\delta}^{(\beta)}(r)\rangle\right\} ,\label{psis}
\end{equation}
where the coefficient $C$ is determined by the condition (\ref{bc2-1})
and the coefficient $B$ could be arbitrary $r$-independent constant.
In addition, according to this result and Eqs. (\ref{psia}, \ref{psib}),
in the region $r>b$ the component $\langle c|\psi_{\epsilon,\delta}(r)\rangle$
is a linear combination of functions $f_{\epsilon-\delta}^{0}(r)$
and $g_{\epsilon-\delta}^{0}(r)$. In our system with $\epsilon<\delta$,
these two functions satisfy \cite{qdt1}
\begin{eqnarray}
\lim_{r\rightarrow\infty}f_{\epsilon-\delta}^{0}(r) & = & \frac{1}{\sqrt{2\pi\kappa}}\left[W_{f-}(\epsilon-\delta)e^{\kappa r}+W_{f+}(\epsilon-\delta)e^{-\kappa r}\right];\nonumber \\
\label{fexp}\\
\lim_{r\rightarrow\infty}g_{\epsilon-\delta}^{0}(r) & = & \frac{1}{\sqrt{2\pi\kappa}}\left[W_{g-}(\epsilon-\delta)e^{\kappa r}+W_{g+}(\epsilon-\delta)e^{-\kappa r}\right],\nonumber \\
\label{gexp}
\end{eqnarray}
where 
\begin{equation}
\kappa=\frac{\sqrt{m(\delta-\epsilon)}}{\hbar}\label{kappa}
\end{equation}
and the funcitons $W_{ij}(z)$ ($i=f,g;$ $j=\pm$) are discussed
in Ref. \cite{qdt1}. Substituting Eqs. (\ref{fexp}, \ref{gexp})
into Eq. (\ref{psis}), we can express $\lim_{r\rightarrow\infty}\langle c|\psi_{\epsilon,\delta}(r)\rangle$
in terms of parameter $C$. Moreover, matching this expression with
the boundary condition (\ref{bc2-1}), we found that
\begin{eqnarray}
C=\frac{K_{co}W_{g-}(\epsilon-\delta)}{W_{f-}(\epsilon-\delta)-K_{cc}W_{g-}(\epsilon-\delta)}.
\end{eqnarray}
Substituting this expression into Eq. (\ref{psis}), we obtain the component
of the scattering wave function $|\psi_{\epsilon,\delta}(r)\rangle$
in the open channel: 
\begin{equation}
\langle o|\psi_{\epsilon,\delta}(r>b)\rangle=\frac{B}{r}\left\{ f_{\epsilon}^{0}(r)-K_{{\rm eff}}\left[\epsilon,\delta\right]g_{\epsilon}^{0}(r)\right\} ,\label{psirb}
\end{equation}
where the function $K_{{\rm eff}}[\epsilon,\delta]$ is defined as
\begin{equation}
K_{{\rm eff}}\left[\epsilon,\delta\right]=K_{oo}^{0}+\frac{K_{oc}^{0}K_{co}^{0}}{\chi(\epsilon-\delta)-K_{cc}^{0}}.\label{keff}
\end{equation}
Here $K_{ij}^{0}$ ($i,j=o,c$) are given in Eqs. (\ref{kcc}, \ref{koc}),
with $K_{\pm}^{0}$ being given in Eq. (\ref{kpm}), and the function
$\chi(z)$ is defined as 
\begin{equation}
\chi(z)=\frac{W_{f-}(z)}{W_{g-}(z)}.\label{chiz}
\end{equation}
Substituting Eqs. (\ref{ff}, \ref{gg}) into Eq. (\ref{psirb}),
we can further obtain the behavior of $\langle o|\psi_{\epsilon,\delta}(r)\rangle$
in the limit $r\rightarrow\infty$. Comparing this expression with
the relation 
\begin{equation}
\lim_{r\rightarrow0}\langle o|\psi_{\epsilon,\delta}(r)\rangle\propto\frac{1}{r}\left[\cot\eta_{0}(k)\sin(kr)+\cos(kr)\right],\label{tane}
\end{equation}
where $\eta_{0}$ is the $s$-wave scattering phase shift, we finally
find that $\tan\eta_{0}(k)$ can be expressed as 
\begin{eqnarray}
\cot\eta_{0}(k)=\frac{Z_{ff}(\epsilon)-K_{{\rm eff}}\left[\epsilon,\delta\right]Z_{gf}(\epsilon)}{K_{{\rm eff}}\left[\epsilon,\delta\right]Z_{gg}(\epsilon)-Z_{fg}(\epsilon)}.\label{taneta}
\end{eqnarray}
Moreover, since $K_{{\rm eff}}\left[\epsilon,\delta\right]$ is determined by the parameters $K_{ij}^{(0)}$ ($i,j=o,c$) and $K_{ij}^{(0)}$
is a function of the scattering lengths $a_{{\rm s}}^{(\pm)}$ and
the characteristic length $\beta_{6}$ of the van der Waals interaction
potential, $\tan\eta_{0}(k)$ given by Eq. (\ref{taneta}) is essentially
a function of $a_{{\rm s}}^{(\pm)}$, $\beta_{6}$, $\delta$ and
$\epsilon$. Here we point out that, the expression (\ref{taneta})
of $\tan\eta_{0}(k)$ has the same form as the one for the case with
a single-channel van der Waals potential (i.e., Eq.(5) of Ref.\cite{qdt1}),
while the parameter $K_{0}$ for the single-channel case should be
replaced by $K_{{\rm eff}}[\epsilon,\delta]$ for our case.

\begin{figure}
\includegraphics[width=6.5cm]{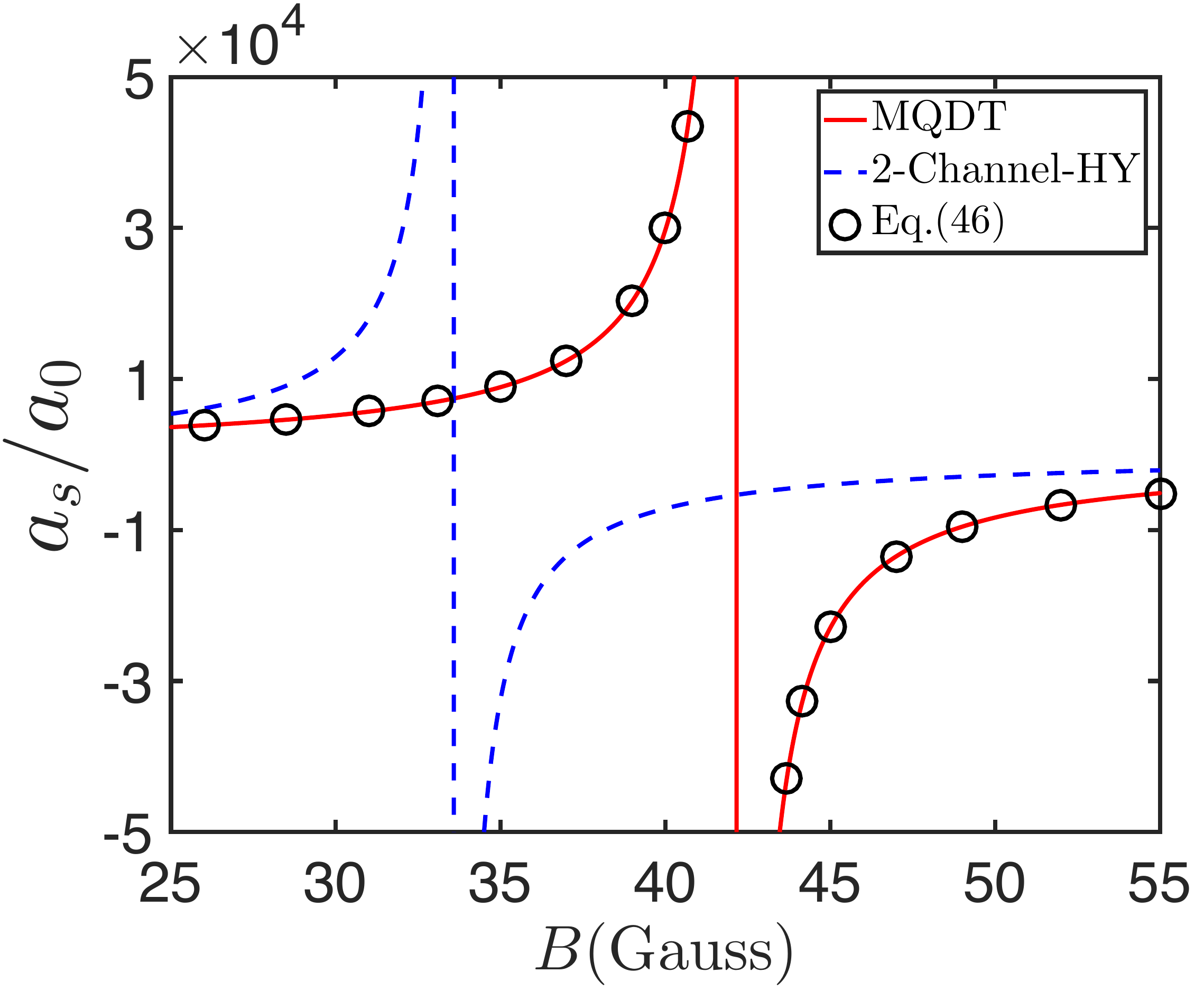} \caption{(color online) The $s$-wave scattering length $a_{s}$ of $^{173}$Yb
atoms, as a function of magnetic field. Here we show the results given
by the MQDT (i.e., Eq. (\ref{as})) (red solid line), the 2-channel
Huang-Yang (HY) pesudopotential (i.e., Eq. (\ref{ashy})) (blue dashed
line) and the result from the approximated expression (\ref{expandas})
(black circles). We consider the case with $\Delta_{m}=5$ and take
$a_{s}^{(+)}=1900a_{0}$, $a_{s}^{(-)}=200a_{0}$, $\beta_{6}=169.6a_{0}$
with $a_{0}$ being the Bohr's radius, and $\mu_{e}-\mu_{g}=2\pi\hbar\times112{\rm Hz}/{\rm Gauss}$.
\label{fig:as}}
\end{figure}

\subsection{$s$-wave Scattering Length and Effective Range}

Using Eq. (\ref{taneta}) we can calculate the two-atom $s$-wave
scattering length $a_{s}$ and effective range $r_{{\rm eff}}$ which
are defined via the low-energy expansion of $k\cot\eta_{0}(k)$: 
\begin{eqnarray}
k\cot\eta_{0}(k)=-\frac{1}{a_{s}}+\frac{1}{2}r_{{\rm eff}}k^{2}+{\cal O}(k^{3}).\label{kcoteta}
\end{eqnarray}
Substituting Eq. (\ref{taneta}) into Eq. (\ref{kcoteta}) and using
direct calculations which are quite similar to the single-channel
case \cite{qdt1}, we can obtain 
\begin{eqnarray}
a_{s}(\delta) & = & \frac{2\pi}{[\Gamma(1/4)]^{2}}\frac{K_{{\rm eff}}[0,\delta]-1}{K_{{\rm eff}}[0,\delta]}\beta_{6},\label{as}\\
r_{{\rm eff}}(\delta) & = & \frac{[\Gamma(1/4)]^{2}}{3\pi}\frac{K_{{\rm eff}}[0,\delta]^{2}+1}{\left(K_{{\rm eff}}[0,\delta]-1\right)^{2}}\beta_{6}\nonumber \\
 &  & +\frac{[\Gamma(1/4)]^{2}}{\pi}\frac{\hbar^{2}K_{{\rm eff}}^{\prime}[0,\delta]}{m\beta_{6}\left(K_{{\rm eff}}[0,\delta]-1\right)^{2}},\label{asreff}
\end{eqnarray}
where $K_{{\rm eff}}^{\prime}[\epsilon,\delta]=dK_{{\rm eff}}^{\prime}[\epsilon,\delta]/d\epsilon$.
In addition, with the help of the relation $\delta=(\delta\mu)B$
we can further express $a_{s}$ and $r_{{\rm eff}}$ as functions
of the magnetic field $B$. It is clear that we have $a_{s}=\infty$
at the magnetic field $B_{0}$ which satisfies the condition 
\begin{eqnarray}
K_{{\rm eff}}[0,(\delta\mu B_{0})]=0.
\end{eqnarray}
That is the OFR.

In Fig.~\ref{fig:as}, we illustrate the scattering length $a_{s}$
for $^{173}$Yb with $a_{s}^{(+)}=1900a_{0},a_{s}^{(-)}=200a_{0}$\cite{ofrexp2}
and $\beta_{6}=169.6a_{0}$\cite{spinexchange3}, with $a_{0}$ being
the Bohr's radius. Here we consider the case with $\Delta_{m}=5$.
For comparison, we also show $a_{s}$ given by the zero-range two-channel
Huang-Yang pseudopotential, which can be expressed as \cite{ourprl,ourpra},
\begin{eqnarray}
a_{s}=\frac{-[a_{s}^{(+)}+a_{s}^{(-)}]+2\sqrt{m\delta/\hbar^{2}}a_{s}^{(+)}a_{s}^{(-)}}{[a_{s}^{(+)}+a_{s}^{(-)}]\sqrt{m\delta/\hbar^{2}}-1}.\label{ashy}
\end{eqnarray}
As show in Fig.~\ref{fig:as}, the difference between the OFR points
given by the MQDT and the 2-channel Huang-Yang pseudopotential is
about 9G, and the relative difference is about 20\%.

On the other hand, around the OFR point $B_{0}$ the scattering length
$a_{s}$ can be expanded as a series of $1/(B-B_{0})$. By neglecting
the high-order terms, we obtain the approximate expression of $a_{s}$
: 
\begin{eqnarray}
a_{s}\approx a_{{\rm bg}}\left(1-\frac{\Delta_{B}}{B-B_{0}}\right).\label{expandas}
\end{eqnarray}
Our calculation show that for $^{173}$Yb we have $a_{{\rm bg}}=-98a_{0}$
and $\Delta_{B}=-660G$. As shown in Fig.~\ref{fig:as}, this approximate
expression is quantitatively consistent with the MQDT result (i.e.,
Eq.(\ref{as})) in a large range of magnetic field.

\begin{figure}
\includegraphics[width=6.5cm]{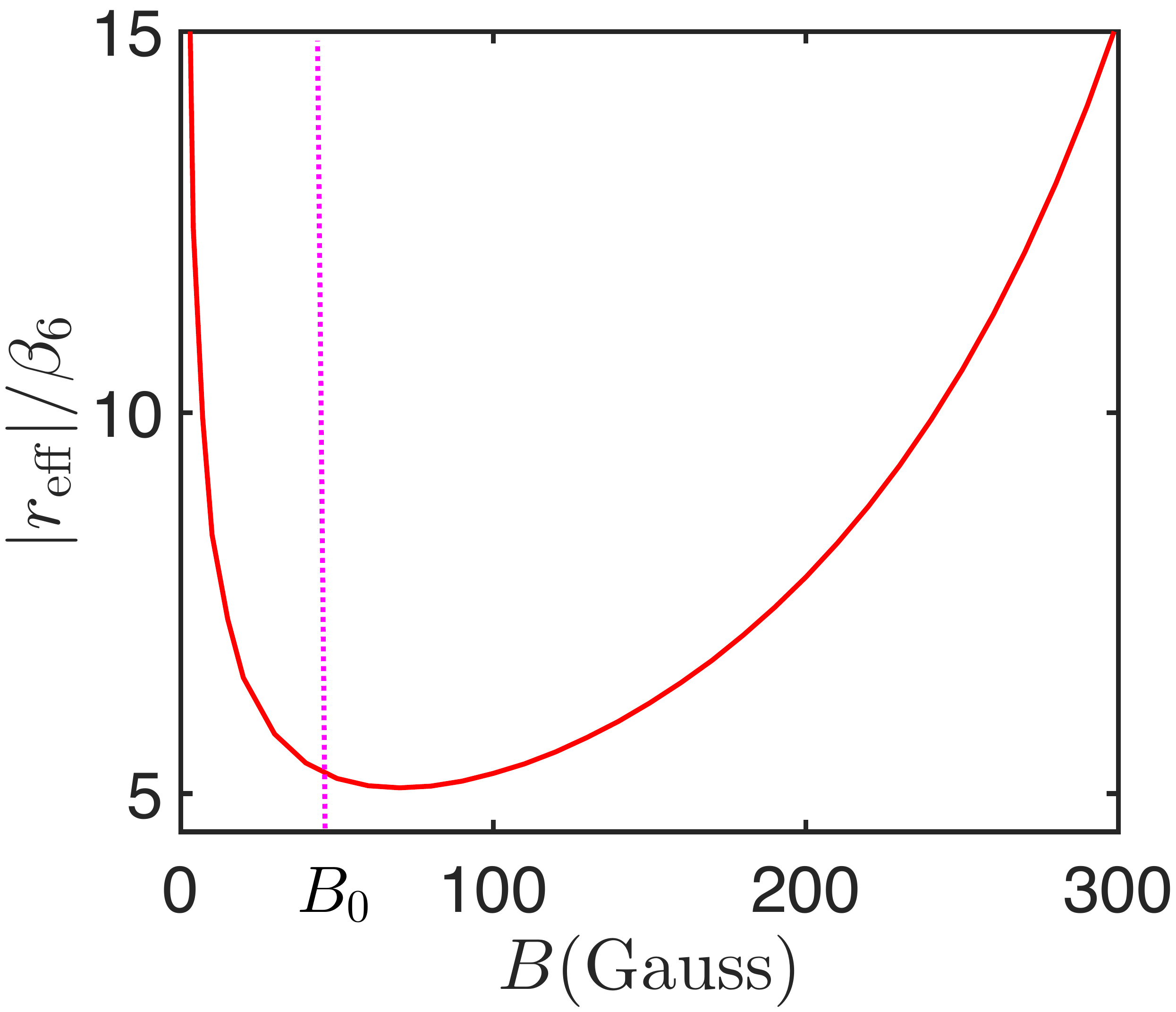} \caption{(color online) The effective range $r_{{\rm eff}}$ of $^{173}$Yb
atoms given by the MQDT. Here the pink dotted line indicates the OFR point $B_0$. In our calculation
we use the same parameter as in Fig.~\ref{fig:as}. \label{fig:reff}}
\end{figure}

In Fig.~\ref{fig:reff} we illustrate the effective range $r_{{\rm eff}}$
for the OFR of $^{173}$Yb atoms with $\Delta_{m}=5$. Our calculation
shows that at the OFR point we have $|r_{{\rm eff}}|\approx908.7a_{0}\approx5.4\beta_{6}$
and thus the resonance parameter $s_{{\rm res}}\equiv4\pi\beta_{6}/(\Gamma(1/4)^{2}r_{{\rm eff}})$ is about $0.18$.
This means that OFR for $^{173}$Yb is a narrow resonance in the sense
of effective range\cite{chinrmp,nr}.

It is pointed that the effective range diverges in the limit $B\rightarrow0$,
as shown in Fig.~\ref{fig:reff}. That is due to the fact that the
function $\frac{d\chi(z)}{dz}|_{z=-\delta}$, which is proportional
to $K^{\prime}[0,\delta]$, diverges in the limit $\delta\rightarrow0$.
This result may also be understood with the following analysis. We
consider the scattering of two atoms incident from the open channel
$|o\rangle$. When the scattering energy $\epsilon$ is smaller than
the energy gap $\delta$ between the open and closed channels, there
is only elastic scattering between these two atoms. Accordingly, the
parameter $k\cot\eta_{0}$ defined by Eq. (\ref{tane}) is real. Nevertheless,
when $\epsilon>\delta$ there are both elastic scattering in channel
$|o\rangle$ and the inelastic scattering from channel $|o\rangle$
to $|c\rangle$. As a result, the imaginary part of $k\cot\eta_{0}$
becomes nonzero. Therefore, as a function of $k$, the factor $k\cot\eta_{0}$
is not analytical at the point $k=\sqrt{\delta}$. Thus, the convergence
radius of the low-energy expansion (\ref{kcoteta}) is at most $\sqrt{\delta}$.
Therefore, in the limit $B\rightarrow0$ which yields $\delta\rightarrow0$,
the convergence radius of (\ref{kcoteta}) tends to zero. As a result,
the expansion coefficient $r_{{\rm eff}}$ diverges.

\section{Two-Atom Bound State}

In this section we investigate the two-atom bound state in the system
with OFR. We will calculate the binding energy and wave function with MQDT and study
the clock-transition spectrum for the bound state, which may be observed
in the experiments.

\subsection{Binding Energy and Wave Function}

In our system the two-body bound state wave function $|\phi_{b}(r)\rangle$
and the bound-state energy $E_{b}$ satisfy the Schr$\ddot{{\rm o}}$dinger
equation 
\begin{eqnarray}
\hat{H}|\phi_{b}(r)\rangle=E_{b}|\phi_{b}(r)\rangle\label{sb}
\end{eqnarray}
as well as the conditions 
\begin{eqnarray}
\lim_{r\rightarrow0}(r|\phi_{b}(r)\rangle) & = & 0;\label{bc1}\\
\lim_{r\rightarrow\infty}|\phi_{b}(r)\rangle & = & 0,\label{bc2}
\end{eqnarray}
and 
\begin{eqnarray}
E_{b}<0.
\end{eqnarray}
Here we consider the cases where the binding energy $|E_{b}|$ is
much smaller than $\hbar^{2}/(m\beta_{6}^{2})$. In these cases we
can derive $E_{b}$ with the MQDT approach introduced above. With
the analysis shown in the above section, we can obtain two special
solutions $|\psi_{E_{b},\delta}^{(\alpha)}(r)\rangle$ and $|\psi_{E_{b},\delta}^{(\beta)}(r)\rangle$
for Eqs. (\ref{sb}) and (\ref{bc1}). In the region $r>b$, the solutions
$|\psi_{E_{b},\delta}^{(\alpha,\beta)}(r)\rangle$ also satisfy Eqs.
(\ref{psia}, \ref{psib}) with $K_{ij}^{0}$ ($i,j=o,c$) being given
by Eqs. (\ref{kcc}, \ref{koc}) and $\epsilon=E_{b}$. The bound-state
wave function $|\phi_{b}(r)\rangle$ can be expressed as the superposition
of these two special solutions, i.e., we have 
\begin{eqnarray}
|\phi_{b}(r)\rangle=C_{\alpha}|\psi_{E_{b},\delta}^{(\alpha)}(r)\rangle+C_{\beta}|\psi_{E_{b},\delta}^{(\beta)}(r)\rangle,\label{phib}
\end{eqnarray}
with $C_{\alpha,\beta}$ the superposition coefficients. Furthermore,
substituting the behaviors of the functions $f_{\epsilon}^{(0)}(r)$
and $g_{\epsilon}^{(0)}(r)$ in the long-range limit $r\rightarrow\infty$,
i.e., Eqs. (2) and (3) of Ref.\cite{qdt1}, we can derive the long-range
behavior of the special solutions $|\psi_{E_{b},\delta}^{(\alpha,\beta)}(r)\rangle$.
Substituting this behavior into the expression (\ref{phib}) and then into
the boundary conditions (\ref{bc1}, \ref{bc2}), we can finally derive
the algebraic equation satisfied by the bound-state energy $E_{b}$
\begin{eqnarray}
\chi(E_{b}) & = & K_{{\rm eff}}[E_{b},\delta],\label{Eb}
\end{eqnarray}
with the function $\chi(z)$ and $K_{{\rm eff}}[z,\delta]$ being
introduced in Sec. II. C. We can obtain the energy $E_{b}$ by solving Eq. (\ref{Eb}). 

\begin{figure}
\includegraphics[width=6.5cm]{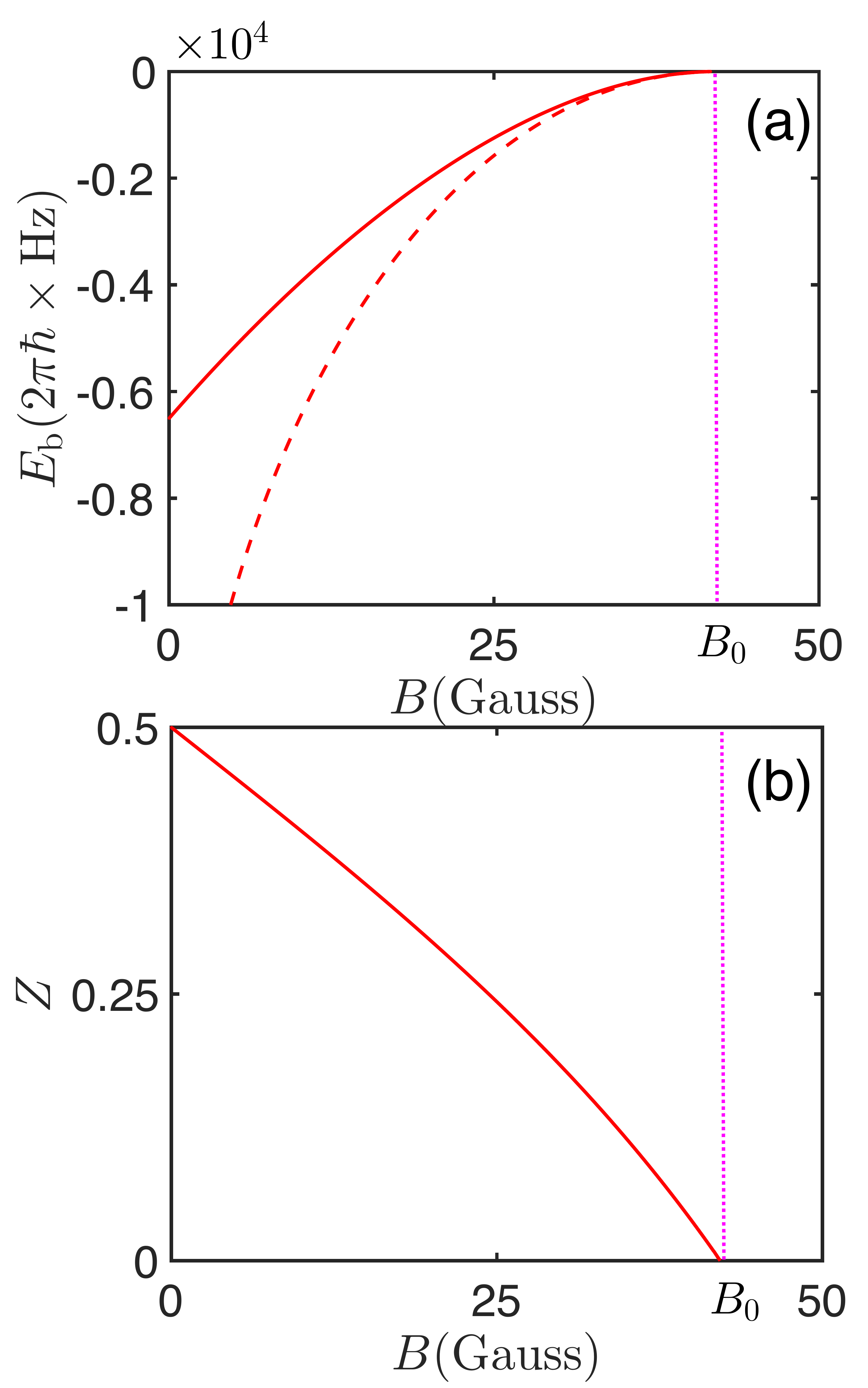} \caption{(color online) The bound-state energy $E_{b}$ (a) and closed-channel
population $Z$ (b) of $^{173}$Yb atoms. Here we show the value of
$E_{b}$ given by the MQDT (red solid line) and the simple expression
$E_{{\rm s}}\equiv-\hbar^{2}/ma_{s}^{2}$ (red dashed line). In our
calculation we use the same parameter as in Fig.~\ref{fig:as}. 
Here the pink dotted line indicates the OFR point $B_0$.
\label{figEb}}
\end{figure}

Furthermore, we can also calculate the closed-channel population
$Z$ of the two-body bound state, which is defined as 
\begin{eqnarray}
Z\equiv\frac{\int|\langle c|\phi_{b}(r)\rangle|^{2}d{\bf r}}{\int\left[|\langle c|\phi_{b}(r)\rangle|^{2}+|\langle o|\phi_{b}(r)\rangle|^{2}\right]d{\bf r}}.
\end{eqnarray}
Using the Feynmann-Hellman theorem, it can be prove that the value
of $Z$ is given by the derivative of the bound-state energy $E_{b}$
with respect to the energy gap $\delta$ between the open and closed
channels: 
\begin{eqnarray}
Z=\frac{\partial E_{b}}{\partial\delta}.\label{FH}
\end{eqnarray}

In Fig.~\ref{figEb} (a, b) we illustrate the bound-state energy
$E_{b}$ and the closed-channel population $Z$ for $^{173}$Yb atoms,
as functions of the magnetic field $B$. For comparison, we also show
the energy given by the the simple expression $E_{{\rm s}}\equiv-\hbar^{2}/ma_{s}^{2}$
with $a_{s}$ being the $s$-wave scattering length given by the MQDT.
For a wide Feshbach resonance which is dominated by the open channel,
we have $E_{b}\approx E_{{\rm s}}$ and $Z\approx0$ in a broad region
around the resonance point. Nevertheless, as shown in Fig.~\ref{figEb}
(a, b), in most of the resonance region of $^{173}$Yb atoms the behaviors
of $E_{b}$ and $E_{{\rm s}}$ are quite different with each other
and the closed-channel population $Z$ is significantly non-zero.
These results also imply that the OFR for $^{173}$Yb is a narrow
resonance in which the contribution from the closed channel is quite
significant. That is consistent with our previous analysis based on
the effective range.

\subsection{Clock-Transition Spectrum}

\begin{figure}
\includegraphics[width=7.0cm]{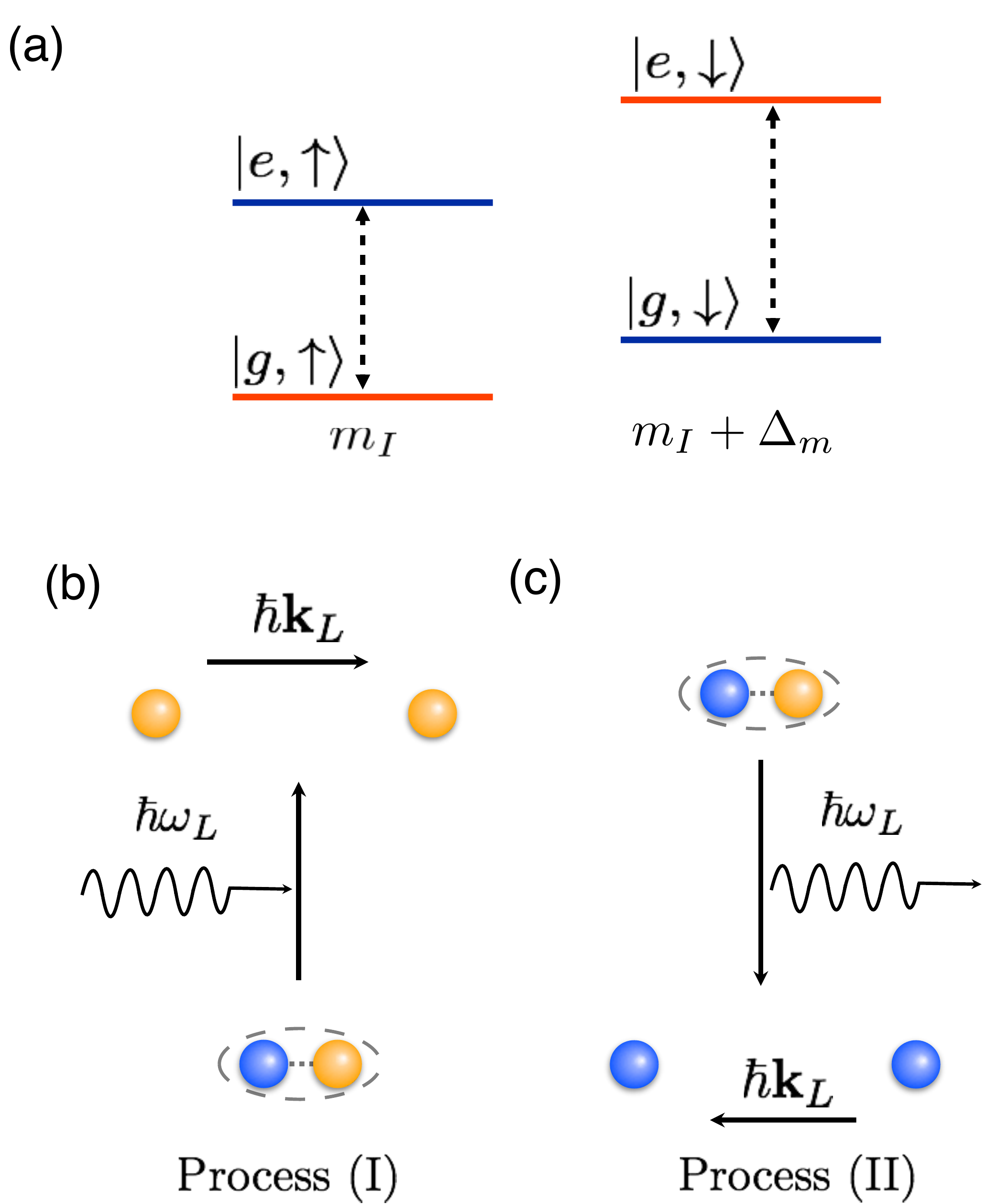} \caption{(color online) 
(a): Schematic diagram for the experiment of clock-transition spectrum.
The $\pi$-polarized clock laser can induce one-body transition between states $|g,\uparrow\rangle$ and $|e,\uparrow\rangle$, as well as the transition between  $|g,\downarrow\rangle$ and $|e,\downarrow\rangle$. As defined in Sec. II, $m_I$ and  $m_I+\Delta_m$ are the magnetic quantum numbers for the nuclear spin states $\uparrow$ and $\downarrow$, respectively.
(b) and (c):
Two processes 
of clock-laser-induced dissociation of a dimer, where one atom (the yellow atom) is in the $e$-state and the other atom (the blue atom) is in the $g$-state. Process (I): the atom in the $g$-state absorbs a photon and transit to the $e$-state. The two-atom mass center gain a photon recoil momentum $\hbar{\bf k}_{\rm L}$.
Process (II): the atom in $e$-state emit a photon and transit to the $g$-state. The two-atom mass center gain a photon recoil momentum $-\hbar{\bf k}_{\rm L}$.
\label{figsketch}}
\end{figure}

Now we investigate the clock-transition spectrum for the ultracold
gases of alkali-earth (like) atoms in the two-body bound state $|\phi_{b}(r)\rangle$
(i.e., the ultracold gases of dimers). It is clear that in each dimer
one atom is in the electronic orbital $g$-state and the other atom
is in the $e$-state. Therefore, if a pulse of clock laser with $\pi$-polarization,
which can induce the one-body transition (clock transition) between
states $|g,j\rangle$ and $|e,j\rangle$ ($j=\uparrow,\downarrow$),
is applied to these two atoms, the dimer may be dissociated into two
free atoms via the following two first-order processes (Fig.~\ref{figsketch}): 
\begin{itemize}
\item[(I)] The atom in $g$-state absorbs a photon and transit to the $e$-state.
After this process both of the two atoms are in the $e$-state. Since in $|\phi_{b}(r)\rangle$ one atom is in nuclear-spin
state $\uparrow$ and other atom is in state $\downarrow$, and
the $\pi$-laser beam does not change nuclear-spin state, after this process we have one atom in state $|e,\uparrow\rangle$
and the other atom in state $|e,\downarrow\rangle$. Furthermore,
the center of mass (CoM) of these two atoms can obtain a recoil momentum $\hbar{\bf k}_{{\rm L}}$ from the laser photon,
with ${\bf k}_{{\rm L}}$ being the wave vector of the clock laser.
 
\item[(II)] The atom in $e$-state emit a photon and transit to the $g$-state.
With similar analysis, we know that after this process one atom is
in state $|g,\uparrow\rangle$ and the other atom is in state $|g,\downarrow\rangle$,
and the CoM can also obtain a recoil momentum
$-\hbar{\bf k}_{{\rm L}}$. 
\end{itemize}

Now we study the properties of the clock-transition spectrum, i.e., the 
dissociation rate as a function of the clock-laser angular frequency $\omega_{\rm L}$.
We first consider the energy condition of the above two processes.
Before the transition process, the energy of these two atoms
is 
\begin{eqnarray}
E_{0}=E_{e\uparrow}+E_{g\downarrow}-|E_{b}|+\frac{\hbar^{2}|{\bf K}|^{2}}{4m},\label{e0}
\end{eqnarray}
where $\hbar{\bf K}$ is the CoM momentum. Here $E_{lj}$
($l=e,g$, $j=\uparrow,\downarrow$) is the energy of the one-atom
internal state $|l,j\rangle$, and can be expressed as $E_{g\uparrow}=\mu_{g}m_{I}B$,
$E_{g\downarrow}=\mu_{g}(m_{I}+\Delta_{m})B$, $E_{e\uparrow}=\epsilon_{eg}+\mu_{e}m_{I}B$
and $E_{e\downarrow}=\epsilon_{eg}+\mu_{e}(m_{I}+\Delta_{m})B$,
with $\epsilon_{eg}$
being the energy gap between $e$-state and $g$-state for $B=0$. 
The term
$E_{e\uparrow}+E_{g\downarrow}$ in Eq. (\ref{e0}) is nothing but
the threshold energy of the open channel $|o\rangle$. Now we consider the process (I) in which the atoms absorb
a photon. Due to the energy-conservation, this process can occur under
the condition 
\begin{eqnarray}
E_{0}+\hbar\omega_{{\rm L}}>E_{{\rm min}}^{{\rm (I)}},\label{se1}
\end{eqnarray}
where $E_{{\rm min}}^{{\rm (I)}}$
is the minimum energy of the finial states of process (I). Furthermore,
since the finial state of process (I) is a scattering state of two
atoms in state $|e,\uparrow\rangle$ and $|e,\downarrow\rangle$,
with mass-center momentum $\hbar({\bf K}+{\bf k}_{{\rm L}})$, the
minimum energy of the finial state of process (I) is 
\begin{eqnarray}
E_{{\rm min}}^{{\rm (I)}}=E_{e\uparrow}+E_{e\downarrow}+\frac{\hbar^{2}|{\bf K}+{\bf k}_{{\rm L}}|^{2}}{4m}.\label{e1}
\end{eqnarray}
Thus, the energy condition (\ref{se1}) for process (I) can be re-written as 
\begin{eqnarray}
\omega_{{\rm L}}>\omega_{{\rm I}}({\bf K})\equiv\frac{E_{e\downarrow}-E_{g\downarrow}+|E_{b}|}{\hbar}+\frac{\hbar\left(|{\bf k}_{{\rm L}}|^{2}+2{\bf K}\cdot{\bf k}_{{\rm L}}\right)}{4m}.\nonumber \\
\label{e1}
\end{eqnarray}
Similarly, since in process (II) the atoms emit a photon, this this
process can occur under the condition 
\begin{eqnarray}
E_{0}-\hbar\omega_{{\rm L}}>E_{{\rm min}}^{{\rm (II)}},\label{se2}
\end{eqnarray}
where $E_{{\rm min}}^{{\rm (II)}}$ is the minimum energy of the finial
states of process (II), and can be expressed as 
\begin{eqnarray}
E_{{\rm min}}^{{\rm (II)}}=E_{g\uparrow}+E_{g\downarrow}+\frac{\hbar^{2}|{\bf K}-{\bf k}_{{\rm L}}|^{2}}{4m}.\label{e2}
\end{eqnarray}
Using this result, we can re-express the energy condition (\ref{se2}) for process
(II) as 
\begin{eqnarray}
\omega_{{\rm L}}<\omega_{{\rm II}}({\bf K})\equiv\frac{E_{e\uparrow}-E_{g\uparrow}-|E_{b}|}{\hbar}-\frac{\hbar\left(|{\bf k}_{{\rm L}}|^{2}-2{\bf K}\cdot{\bf k}_{{\rm L}}\right)}{4m}.\nonumber \\
\label{e2}
\end{eqnarray}
The above analysis yields that the laser-induced dissociation process
can only occur under the condition (\ref{e1}) or (\ref{e2}). 
In
particular, the dissociation process cannot occur in the frequency
region $\omega_{{\rm II}}<\omega<\omega_{{\rm I}}$. Thus, the clock-transition spectrum includes two branches corresponding to process (I) and (II), respectively.

Our above analysis can be verified by the quantitative calculation
for the dissociation rate based on the Fermi's golden rule. We consider
two atoms with initial wave function 
\begin{eqnarray}
|\Psi({\bf R},{\bf r},0)\rangle=\frac{1}{(2\pi)^{\frac{3}{2}}}\int d{\bf K}\phi({\bf K})e^{i{\bf K}\cdot{\bf R}}|\phi_{b}(r)\rangle,\label{psi0}
\end{eqnarray}
where ${\bf R}$ and ${\bf r}$ are the mass-center position and the
relative position of these two atoms, respectively, $|\phi_{b}(r)\rangle$
is the two-atom bound sate wave function we obtained in the above
subsection, and $\phi({\bf K})$ is the wave function of the CoM
motion in the momentum space. We further assume that the laser beam
is applied from the time $t=0$.

At time $t$ the probability of the two atoms being in the bound state can
be denoted as $P(t)$. The Fermi's golden rule yields that (Appendix
B) when $t$ is short we have  \cite{FG}
\begin{eqnarray}
P(t)\approx1-\Gamma t.\label{bigpt}
\end{eqnarray}
Here $\Gamma$ is the dissociation rate. Furthermore, as shown in Appendix B
for our system it can be proved that 
\begin{eqnarray}
\Gamma=\int d{\bf K}|\phi({\bf K})|^{2}\gamma({\bf K}),\label{gamma}
\end{eqnarray}
where $\gamma({\bf K})$ is the dissociation rate corresponding to
the mass-center momentum $\hbar{\bf K}$, and can be expressed as
\begin{eqnarray}
\gamma({\bf K})=\gamma_{{\rm I}}({\bf K})+\gamma_{{\rm II}}({\bf K}).\label{g1}
\end{eqnarray}
Here $\gamma_{l}({\bf K})$ ($l={\rm I},{\rm II}$) is the dissociate
rate for process ($l$), and is given by 
\begin{eqnarray}
\gamma_{l}({\bf K})=\frac{2\pi}{\hbar}\sum_{j=1,2} &  & \int d{\bf k}\left|\int d{\bf r}\langle\Psi_{j}^{l}({\bf k},{\bf r})|\Lambda({\bf r})|\phi_{b}(r)\rangle\right|^{2}\nonumber \\
 &  & \times\delta\left(\hbar\omega_{l}({\bf K})+\xi_{l}\frac{\hbar^{2}|{\bf k}|^{2}}{2m}-\hbar\omega_{{\rm L}}\right),\nonumber\\\label{gam}
\end{eqnarray}
with $\xi_{{\rm I}}=1$, $\xi_{{\rm II}}=-1$ and $\omega_{{\rm I,II}}({\bf K})$
being defined in Eqs (\ref{e1}, \ref{e2}). Here the operator $\Lambda({\bf r})$
being defined as $\Lambda({\bf r})=\frac{\hbar\Omega}{2}(|e\rangle^{(1)}\langle g|e^{i{\bf k}_{\rm L}\cdot{\bf r}/2}+|e\rangle^{(2)}\langle g|e^{-i{\bf k}_{\rm L}\cdot{\bf r}/2}+h.c.$,
where $\Omega$ is the Rabi frequency of the laser and $|e(g)\rangle^{(i)}$
($i=1,2$) denotes the electronic orbital state of the $i$-th atom.
In Eq. (\ref{gam}) $|\Psi_{j}^{l}({\bf k},{\bf r})\rangle$ ($l={\rm I},{\rm II}$,
$j=1,2$) is the finial state of process $({l})$, i.e., the scattering
state of two atoms with incident momentum ${\bf k}$ and two-atom
electronic orbital state $|l\rangle$ and two-atom nuclear-spin state
$|j\rangle$, which are defined as $|{\rm I}\rangle=|ee\rangle$,
$|{\rm II}\rangle=|gg\rangle$, $|1\rangle=(|\uparrow\downarrow\rangle+|\downarrow\uparrow\rangle)/\sqrt{2}$
and $|2\rangle=(|\uparrow\downarrow\rangle-|\downarrow\uparrow\rangle)/\sqrt{2}$.

\begin{figure}
\includegraphics[width=7.0cm]{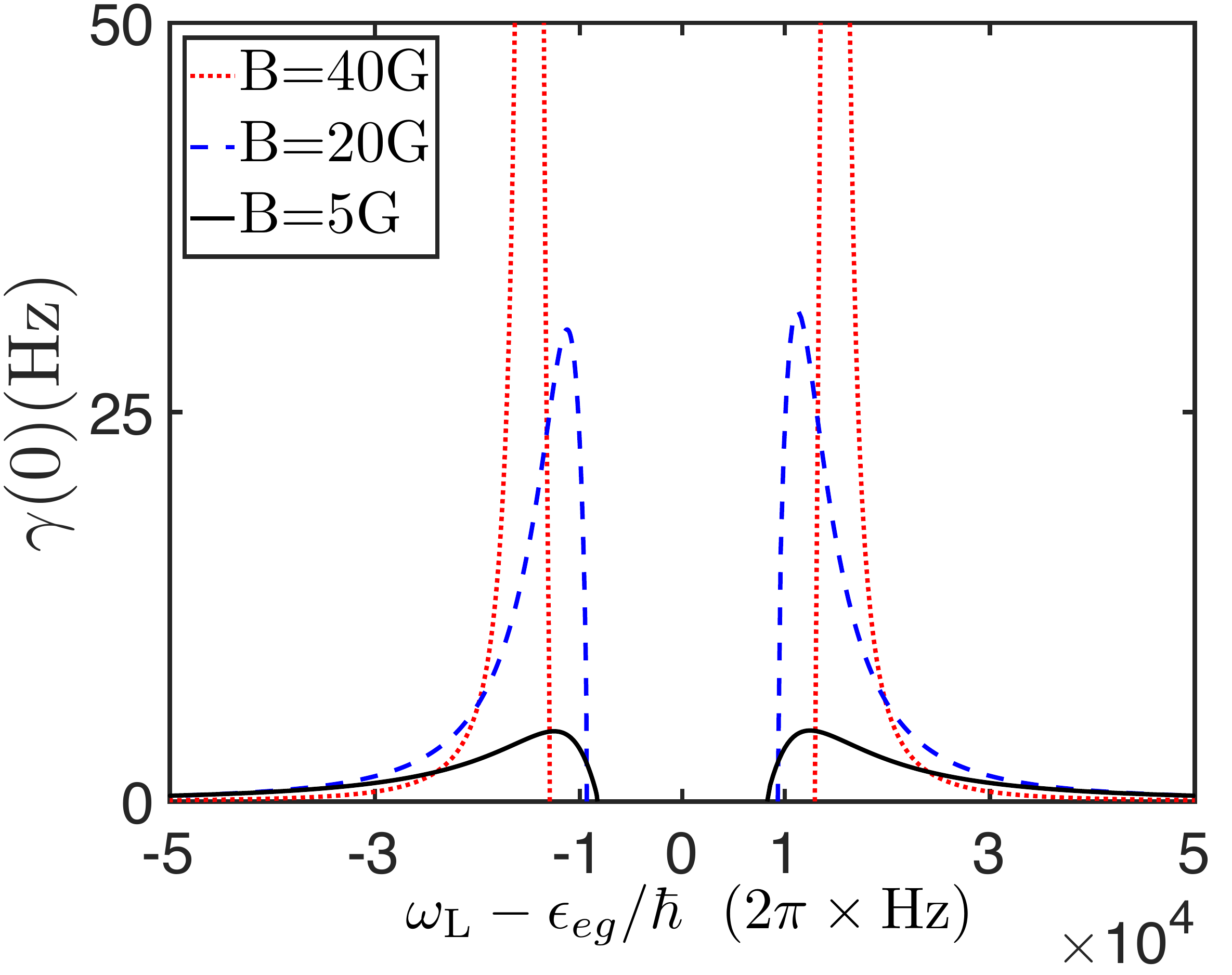} 
\caption{The clock-transition spectrum for two $^{173}$Yb atoms
in bound state $|\phi_b(r)\rangle$ with $B=5$G ($|E_b|=\hbar\times(2\pi)5176.05$Hz, black solid line), $20$G ($|E_b|=\hbar\times(2\pi)2007.59$Hz, blue dashed line)
and $40$G ($|E_b|=\hbar\times(2\pi)22.63$Hz, red dotted line).
In our calculation we take $m_I=-5/2$, $\Delta_m=5$, $\Omega=(2\pi)10^3$Hz, $\epsilon_{eg}=\hbar\times(2\pi)5.19\times10^{14}{\rm Hz}$ \cite{LENS-se, Bloch-se}, $a_{gg} = 199.4a_0$, $a_{ee} = 306.2a_0$ \cite{spinexchange3}, and $\mu_e-\mu_g=\hbar\times(2\pi){\rm 112}$Hz/Gauss.
The maximum value of $\gamma({\bf 0})$ for $B=40$G is $772$Hz. 
For a given magnetic field, the right and left branch of the spectrum
corresponds to transition processes (I) and (II), respectively.
\label{figspectrum}}
\end{figure}

Furthermore, Eq. (\ref{gamma}) implies that if the mass-center momentum
is mainly distributed in a small region around a central momentum
${\bf K}_{0}$, we have $\Gamma\approx\gamma({\bf K}_{0})$. In the
following we consider the simple case with ${\bf K}_{0}={\bf 0}$.
We further calculate $\gamma({\bf 0})$ for $^{{\rm 173}}$Yb atoms
for the cases with different magnetic field. Our calculation is based
on the binding energy $E_{b}$ and the closed-channel population $Z$
derived with the MQDT in Sec. III. A. On the other hand, since in
our system both $E_{b}$ and the energy gap $\delta$ between the
open and closed channels are much smaller than the van der Waals energy,
in the bound state $|\phi_{b}(r)\rangle$ the two-atom relative position
$r$ is mainly distributed in the region $r\gtrsim\beta_{6}$. Thus,
in our calculation we ignore the contribution
from the two-atom relative function $|\phi_{b}(r)\rangle$ in the
region $r\lesssim\beta_{6}$ and use the approximated bound-state wave function
\begin{equation}
|\phi_{b}(r)\rangle=\sqrt{1-Z}\frac{e^{-r/r_{o}}}{\sqrt{2\pi r_o} r}|o\rangle+\sqrt{Z}\frac{e^{-r/r_{c}}}{\sqrt{2\pi r_c}r}|c\rangle,
\end{equation}
and  the approximated scattering-state wave functions
\begin{eqnarray}
&&|\Psi_{j}^{l}({\bf k},{\bf r})\rangle = \frac{1}{(2\pi)^{\frac{3}{2}}}\left[e^{i{\bf k}\cdot{\bf r}}+\frac{-1}{ik+\frac{1}{a_{l}}}e^{ikr}\right]|l\rangle|j\rangle;\nonumber\\
&&{\rm for}\ l={\rm I},\ {\rm II}\ {\rm and}\ j=1,2. 
\end{eqnarray}
Here $r_{o}=\hbar/\sqrt{2m|E_{b}|}$, $r_{c}=\hbar/\sqrt{2m(|E_{b}|+\delta)}$, $a_{\rm I}=a_{ee}$, and $a_{\rm II}=a_{gg}$, with $a_{{ee}}$ ($a_{{gg}}$) being the scattering length of
two atoms who are both in $e$-state ($g$-state).

In Fig.(\ref{figspectrum}) we show 
$\gamma({\bf 0})$ as a function of $\omega_{\rm L}$
for $^{{\rm 173}}$Yb atoms with various magnetic field. It is clear
that for each magnetic field the clock-transition spectrum has
two branches, corresponding to process (I) (right branch with $\omega_{{\rm L}}>\omega_{{\rm I}}$)
and process (II) (left branch with $\omega_{{\rm L}}<\omega_{{\rm II}}$), as we have
analyzed  before.
Furthermore, it is also shown that when the magnetic field is close
to the OFR point $B_{0}$, the spectrum becomes more sharp. This result
may be explained as follows. When the system is close to the OFR point,
the wave packet of the bound state becomes very wide in the real space,
and thus very narrow in the momentum space. Therefore, in this case the bound
state has significant overlap (the Frank-Condon factor) only with
the the scattering states $|\Psi_{j}^{l}({\bf k},{\bf r})\rangle$
with incident momentum and scattering energy being in a small region, and thus
the transition spectrum becomes narrow.

Our above results, together with Eqs. (\ref{e1}, \ref{e2}), show
that both the position and the shape of the clock-transition spectrum
are related to the binding energy $E_{b}$ and the wave function of the two-body bound state
$|\phi_{b}(r)\rangle$. Thus, in the experiments one can detect the properties of $|\phi_{b}(r)\rangle$ via the clock-transition spectrum.

\section{Summary}

In this paper we solve the two-body problem of two alkali-earth (like)
atoms with OFR with the approach of MQDT, in which the effect induced
by the van der Waals interaction potential can be analytically included.
We derive the analytical expression of the scattering length (Eq.
(\ref{as})) and the effective range (Eq. (\ref{asreff})), as well
as the algebraic equation (\ref{Eb}) for the binding energy of the
two-body bound state. We further investigate the clock-transition
spectrum for our system, which can be used for the experimental detection
of the bound state. Since the MQDT approach is quantitatively applicable
for the system where the all the characteristic energies are much
smaller than the van der Waals energy, e.g., the $^{173}$Yb atoms
near the OFR point, our results are helpful for both theoretical and
experimental research for these systems.
\begin{acknowledgments}
We thank Bo Gao and Hui Zhai for helpful discussions. This work has been supported
by the Natural Science Foundation of China under Grant No. 11434011,
and by NKBRSF of China under Grant No. 2012CB922104, the Fundamental
Research Funds for the Central Universities, and the Research Funds
of Renmin University of China under Grant No.16XNLQ03. 
\end{acknowledgments}

\addcontentsline{toc}{section}{Appendices}\markboth{APPENDICES}{}

\appendix

\section{The $\epsilon$- and $\delta$-dependence of $K_{ij}^{0}$}

In this appendix we show that the parameters $K_{ij}^{0}$ in Sec.
II. B are independent of $\epsilon$ and $\delta$. This fact can
be understood with the following arguments.

First, the conditions (\ref{b}, \ref{b2}) imply that there is a
spatial region in which both of the two conditions $r>b$ and $\hbar^{2}\beta_{6}^{4}/(mr^{6})>>\hbar^{2}/(m\beta_{6}^{2})$
are satisfied. Here we denote this region as ${\cal R}$.

Second, similar as in Sec. II. A, due to the conditions (\ref{b},
\ref{b2}) and the fact that both $\epsilon$ and $\delta$ are much
smaller than $\hbar^{2}/(m\beta_{6}^{2})$, in both region 
${\cal R}$ and the region of $r<b$, these two energies
can be neglected from the Schr$\ddot{{\rm o}}$dinger equation (\ref{se}).
Thus, we know that Eq. (\ref{se}) and the condition (\ref{bc}) have
two special solutions which are independent of $\epsilon$ and $\delta$
in the region of ${\cal R}$ and the region of
$r<b$. We denote these two solutions as $|\Phi_{1}(r)\rangle$ and
$|\Phi_{2}(r)\rangle$.

Third, since both $|\psi_{\epsilon,\delta}^{(\alpha,\beta)}(r)\rangle$
introduced in Sec. II. B and $|\Phi_{1,2}(r)\rangle$ are special
solutions of Eqs. (\ref{se}, \ref{bc}), $|\psi_{\epsilon,\delta}^{(\alpha,\beta)}(r)\rangle$
can be expressed as linear superpositions of $|\Phi_{1,2}(r)\rangle$.
Namely, we have 
\begin{eqnarray}
|\psi_{\epsilon,\delta}^{(l)}(r)\rangle=\sum_{j=1,2}A_{j}^{(l)}|\Phi_{j}(r)\rangle.\label{a1}
\end{eqnarray}
Here the coefficients $A_{j}^{(l)}$ ($l=\alpha,\beta$, $j=1,2$)
are determined by the following two conditions given by Eqs. (\ref{psia}, \ref{psib}):
(a) For $r\in {\cal R}$, if one expand $r\langle c|\psi^{(\alpha)}_{\epsilon,\delta}(r)\rangle$
as a superposition of the functions $f_{\epsilon-\delta}^0(r)$ and $g_{\epsilon-\delta}^0(r)$,
then the coefficient for $f_{\epsilon-\delta}^0(r)$  is zero. (b) For $r\in {\cal R}$, if one expand $r\langle o|\psi^{(\alpha)}_{\epsilon,\delta}(r)\rangle$
as a superposition of the functions $f_{\epsilon}^0(r)$ and $g_{\epsilon}^0(r)$,
then the coefficient for $g_{\epsilon}^0(r)$  is zero.
Since these two conditions are independent of $K_{ij}^{0}$
the value of $A_{j}^{(l)}$
($l=\alpha,\beta$, $j=1,2$) is also independent of $K_{ij}^{0}$.
Furthermore, since  in the region ${\cal R}$ both $|\Phi_{1,2}(r)\rangle$
and
the
functions $f_{\epsilon}^{0}(r)$, $g_{\epsilon}^{0}(r)$, $f_{\epsilon-\delta}^{0}(r)$
and $g_{\epsilon-\delta}^{0}(r)$ are independent of $\epsilon$ and
$\delta$ \cite{van}, it is clear that the coefficients $A_{j}^{(l)}$ determined by the above two conditions are also $(\epsilon,\delta)$-independent.
Therefore, the right-hand side of Eq. (\ref{a1})
is  independent of $\epsilon$ and
$\delta$ for $r\in{\cal R}$.
Using this result and Eqs. (\ref{psia},
\ref{psib}), we immediately know that $K_{ij}^{0}$ in Sec. II. B
are independent of $\epsilon$ and $\delta$.

\section{Calculation of dissociation rate }

In this appendix we calculate the dissociation rate of two-atom
bound state and prove Eqs. (\ref{gamma}, \ref{g1}, \ref{gam}).  
In the Schr$\ddot{{\rm o}}$dinger picture, the Hamiltonian for our
problem is given by
\begin{equation}
H=\frac{-\hbar^2\nabla_{\bf R}^{2}}{4m}+H_{I}+H_{{\rm rel}}+H_{{\rm L}}\label{hs}
\end{equation}
where ${\bf R}$ is the two-atom center of mass (CoM) position. Here $H_{I}$
describeds the one-body internal-state energy and is given by
\begin{equation}
H_{I}=\sum_{j=1,2}\sum_{l=e,g}\sum_{s=\uparrow,\downarrow}E_{ls}
|l,s\rangle^{(j)}\langle l,s|,\label{hi}
\end{equation}
with $|l,s\rangle^{(j)}$ ($j=1,2$; $l=e,g$; $s=\uparrow,\downarrow$)  being the internal state of the $s$-th atom and $E_{lj}$ being the corresponding one-body energy, which is
defined in Sec. III.B. In Eq. (\ref{hs}) $H_{{\rm rel}}$ and $H_{\rm L}$ are the
Hamiltonian for the two-atom relative motion and laser-atom coupling,
respectively, and can be expressed as
\begin{eqnarray}
H_{{\rm rel}} & = & \frac{-\nabla_{\bf r}^2}{m}+U_T({\bf r});\label{hrel}\\
H_{{\rm L}} & = & \frac{\hbar\Omega}{2}\sum_{j=1,2}|e\rangle^{(j)}\langle g|e^{i({\bf k}_{\rm L}\cdot{\bf r}_{j}-\omega_{\rm L}t)}+h.c..\label{hl}
\end{eqnarray}
Here ${\bf r}$ is the two-atom relative position, $|e(g)\rangle^{(j)}$ ($j=1,2$) is the electronic-orbital state of the $j$-th atom, $U_T({\bf r})$ is the total interaction potential,
$\Omega$ is the Rabi frequency of the clock laser beam, and ${\bf k}_{\rm L}$
and $\omega_{\rm L}$ being the wave vector and angular frequency of this laser
beam, respectively. In Eq. (\ref{hl}) ${\bf r}_{j}$ ($j=1,2$) is the position
of the $j$-th atom, and can be expressed as
\begin{eqnarray}
{\bf r}_{1} & = & {\bf R}+\frac{{\bf r}}{2};\label{r1}\\
{\bf r}_{2} & = & {\bf R}-\frac{{\bf r}}{2}.\label{r2}
\end{eqnarray}

As shown in Eq. (\ref{psi0}) of Sec. III.B, 
we assume that at $t=0$ 
the two-atom initial wave function is
\begin{eqnarray}
|\Psi({\bf R},{\bf r},t=0)\rangle=\frac{1}{(2\pi)^{\frac{3}{2}}}\int d{\bf K}\phi({\bf K})e^{i{\bf K}\cdot{\bf R}}|\phi_{b}(r)\rangle,\label{psi0a}
\end{eqnarray}
where $|\phi_b(r)\rangle$ is the wave function of the two-atom bound state.
We further assume that the laser beam is applied from $t=0$. Thus,
for $t\geq0$ the evolution of the two atoms is governed by the total
Hamiltonian $H$. At time $t$ the two-atom wave function can be denoted as
$|\Psi({\bf R},{\bf r},t)\rangle$, and the probability of these two atoms being
at the bound state $|\phi_{b}(r)\rangle$ can be expressed
as
\begin{equation}
P(t)=\int d{\bf R}\int d{\bf r}\left|\langle\phi_b(r)|\Psi({\bf R},{\bf r},t)\rangle\right|^2.
\label{pt}
\end{equation}

To calculate $P(t)$ with the Fermi's golden rule, we introduce a
unitary transformation
\begin{equation}
{\cal U}=e^{-i{\bf k}_{L}\cdot{\bf R}\Sigma_{e}}\label{calu}
\end{equation}
where
\begin{equation}
\Sigma_{e}=|ee\rangle\langle ee|-|gg\rangle\langle gg|.\label{ne}
\end{equation}
We further define the wave function $|\Phi({\bf R}, {\bf r}, t)\rangle$ as 
\begin{equation}
|\Phi({\bf R}, {\bf r}, t)\rangle={\cal U}|\Psi({\bf R}, {\bf r}, t)\rangle,\label{bigphit}
\end{equation}
i.e., $|\Phi({\bf R}, {\bf r}, t)\rangle$ is the two-atom state in the rotated frame
induced by ${\cal U}$. It is easy to prove that we have
\begin{eqnarray}
|\Phi({\bf R}, {\bf r}, t=0)\rangle & = & |\Psi({\bf R}, {\bf r}, t=0)\rangle;\label{rel0}\\
P(t)&=&\int d{\bf R}\int d{\bf r}\left|\langle\phi_b(r)|\Phi({\bf R},{\bf r},t)\rangle\right|^2.
\label{pt2}
\nonumber\\
\label{relpt}
\end{eqnarray}
Furthermore, we can also prove that $|\Phi({\bf R},{\bf r},t)\rangle$ satisfies the
Schr${\rm \ddot{o}}$dinger equation
\begin{equation}
i\hbar\frac{d}{dt}|\Phi({\bf R},{\bf r},t)\rangle=H_{\rm rot}|\Phi({\bf R},{\bf r},t)\rangle\label{rotse}
\end{equation}
with $H_{{\rm rot}}$ being the Hamiltonian in the rotated frame
and can be expressed as
\begin{eqnarray}
H_{\rm rot} & = & 
\frac{(-i\hbar\nabla_{\bf R}+\hbar{\bf k}_{\rm L}\Sigma_e)^2}{4m}
+H_I+H_{\rm rel}+h_{\rm L},\nonumber\\
\label{hrot}
\end{eqnarray}
with
\begin{eqnarray}
h_{\rm L}&=&\frac{\hbar\Omega}{2}e^{-i\omega_{\rm L}t}\left(|e\rangle^{(1)}\langle g|e^{i{\bf k}_{\rm L}\cdot{\bf r}/2}+|e\rangle^{(2)}\langle g|e^{-i{\bf k}_{\rm L}\cdot{\bf r}/2}\right)\nonumber\\
&&+h.c..
\end{eqnarray}
Eq. (\ref{hrot}) shows that in the rotated frame the momentum of the CoM is conserved. Using this fact and Eqs. (\ref{rel0}) and (\ref{psi0a}), we can simplify the calculation of the probability $P(t)$ in Eq. (\ref{pt2}) and obtain
\begin{equation}
P(t)=\int d{\bf K}|\phi({\bf K})|^{2}p({\bf K})\label{pt}
\end{equation}
where $p({\bf K})$ is given by
\begin{equation}
p({\bf K})=\int d{\bf r}\left|\langle\phi_{{\bf K}}({\bf r},t)|\phi_{b}(r)\rangle\right|^{2}.\label{spk}
\end{equation}
Here the wave function $|\phi_{{\bf K}}({\bf r},t)\rangle$ satisfies the 
equation
\begin{eqnarray}
i\frac{d}{dt}|\phi_{{\bf K}}({\bf r},t)\rangle=h({\bf K})|\phi_{{\bf K}}({\bf r},t)\rangle
\end{eqnarray}
with
\begin{eqnarray}
h({\bf K}) & = & \frac{-\hbar^2\nabla_{\rm r}^2}{m}+H_{I}+U({\bf r})
+\frac{\left(\hbar{\bf K}+\hbar{\bf k}_{\rm L}\Sigma_e\right)^2}{4m}
+h_{\rm L}.\nonumber\\\label{smallhk}\\
 & \equiv & h_{0}({\bf K})+h_{\rm L},
\end{eqnarray}
as well as the initial condition 
\begin{eqnarray}
|\phi_{{\bf K}}({\bf r},t=0)\rangle=|\phi_b(r)\rangle.\label{ip}
\end{eqnarray}
Eqs. (\ref{spk}-\ref{ip}) show that
to calculate $p({\bf K})$ we need to solve a quantum
evolution problem  governed
by the Hamiltonian $h({\bf K})$. 
This problem is defined in the Hilbert space ${\cal H}_{r}\otimes{\cal H}_I$,
with ${\cal H}_{r}$ and ${\cal H}_I$ being the space for two-atom spatial relative motion and two-atom internal state, respectively, and the CoM momentum $\hbar{\bf K}$
just behaves as a classical parameter (c-number).
In this problem, the term $h_{\rm L}$
induces the transitions from the isolated state $|\phi_{b}(r)\rangle$,
which is a discrete eigen-state of $h_{0}({\bf K})$, to other continuous
eigen-states of $h_{0}({\bf K})$, i.e., the scattering states of
two atoms in either the electronic-orbital state $|ee\rangle$ or $|gg\rangle$. 
Thus, we can calculate $p({\bf K})$ using the Fermi's golden
rule and obtain that when the time $t$ is small enough we have
\cite{FG}
\begin{equation}
p({\bf K})=1-\gamma({\bf K})t\label{pp2}
\end{equation}
where $\gamma({\bf K})$ is given by
Eqs. (\ref{g1}, \ref{gam})
in Sec. III. B. Furthermore, substituting Eq. (\ref{pp2})
into Eq. (\ref{pt}) and using Eq. (\ref{bigpt}) in Sec. III. B,
we can obtain Eq. (\ref{gamma}) in Sec. III. B.


\begin{thebibliography}{10}
\bibitem{FG} Precisely speaking, the Fermi's
golden rule is applicable when both of the following two conditions are satisfied:
(a) $t$ is short enough so that the atom-laser coupling can be considered as a lowest-order perturbation.
(b) $t$ is much longer than the correlation time of the laser-induced
coupling between the bound state and the scattering states.
\end{thebibliography}
\end{document}